# Accountable yet Anonymous AI Agents: Split-Knowledge Binding in China’s National Agent-Identity Layer

**Yifan He[1]*, Zhiguang Shan[2], Le Luo[3], Wei Wang[4]**

[1] Red Date Technology (Hong Kong) Limited, Hong Kong, China

[2] State Information Center (SIC), Beijing, China

[3] China Organization Data Service (CODS), Beijing, China

[4] China Internet Network Information Center (CNNIC), Beijing, China

* Corresponding author: yifan.he@reddatetech.com

---

**Preprint notice (v1, pre-launch).** This version presents the conceptual framework, architecture, threat model, and governance analysis of a national system that is built and scheduled for public launch in Q3 2026 — a framework-and-architecture disclosure, not a deployment study. It reports no deployment measurements by design: Section 5 pre-registers the questions and disclosure protocol under which a later version (v2) will report them after launch.

---

## Abstract

The emerging infrastructure for AI-agent identity has converged, in industry practice and in research proposals alike, on a single resolution of the tension between accountability and privacy: make every agent identifiable. We document a national system in China — built as national infrastructure and scheduled for public launch in Q3 2026 — that occupies a different and underexplored point in the same design space: an agent is *associated* with a verified legal principal without that principal being *disclosed* to any business-layer participant. Re-identification is possible only to a legal authority acting through due process, and only by separately compelling two organizationally distinct government agencies (MPS and SIC), neither of which can re-identify alone. We name the mechanism *split-knowledge binding* and are candid that it is conditional: the separation is structural and procedural, not cryptographic, and a state empowered to compel both agencies can re-identify.

The paper makes five conceptual contributions. (1) **Split-knowledge binding** — a coined mechanism for escrowed accountability, differentiated from prior cryptographic constructions by its institutional rather than cryptographic separation. (2) **The ex-post attribution thesis** — the argued claim that for AI agent actions with legal consequences, only attribution-based accountability carries legal force. (3) **The accountability surface** — a design concept identifying which agent actions leave identity-bearing traces, with a mapped granularity spectrum. (4) **A proportionality framework for identity escrow** — a decision structure grounded in legal proportionality [Barak 2012; Alexy 2002] that selects among three trust architectures (user-sovereign, state-escrow, cryptographic-threshold) and predicts agent-economy fragmentation at the rule-of-law boundary. (5) **The reflexive jurisdiction method** — an evaluative standard that applies the framework's rule-of-law condition to the deployment jurisdiction, distinguishes what the architecture determines from what the jurisdiction determines, and is administered to the paper's own deployment without adjudicating the result, which is a jurisdiction-specific legal question outside this paper's scope. The system is evidence of feasibility at national scale; the framework is the instrument by which any deployment — including this one — should be judged.

---

## 1. Introduction

The infrastructure now being built to give AI agents identities has, with striking uniformity, answered one of the oldest questions in computer security — *must accountability be purchased with identifiability?* — in the affirmative. Industry agent-identity practice treats "no anonymous agents" as a baseline requirement: leading control-plane vendors register every agent, bind it cryptographically to a verified human or organizational sponsor, and maintain continuous operator-side auditability of what each agent does [Okta 2025; Ping Identity 2025; Descope 2025; authID 2025]. The premise is stated almost as a slogan — that identity is becoming "the universal language of accountability, for humans and agents alike" [Ping Identity 2025]. Research proposals locate accountability in the same place. Verifiable user–code binding schemes attest who deployed which code [BAID 2025; South et al. 2025], and "Know Your Agent" attestation frameworks build public, reputation-bearing agent identities so that counterparties can make trust decisions [Birch and Hoffart 2025]. Whether the disclosure runs to the operating platform, to counterparties, or to the public, accountability is made to rest on the revelation of identity.

We want to be precise about what is and is not novel in observing this. The tension between accountability and privacy is not unexamined; it is well recognized in the agent-governance literature, which has carefully articulated the privacy costs of identifiability [Chan et al. 2024]. What is largely absent is exploration, *at infrastructure scale*, of designs that preserve accountability while disclosing identity to no business-layer party at all — reserving disclosure for a lawfully authorized authority acting under due process. That is the region of the design space this paper occupies. We do not claim the privacy/accountability tension was unnoticed, nor that this is the first work to see the privacy costs of identifiability, nor that anything here is a paradigm shift. We claim something narrower and, we think, more useful: that there is an underexplored point in the design space of agent accountability, and that it has been built as national infrastructure.

## 1.1 Why agents, why now

The case for attending to this now rests not on market euphoria but on a structural property of agents. Agents act semi-autonomously, are cheap to create, and cross organizational boundaries, which makes the question "who is responsible?" structurally harder to answer than it is for traditional software [Chan et al. 2023]. Responsibility in multi-agent settings cannot be reduced to local alignment of individual components; it requires systemic governance that can attribute outcomes across a distributed, fast-moving population of actors [Constantinescu and Kaptein 2025]. Enterprise adoption is accelerating — by one widely cited industry projection, some 40% of enterprise applications will feature task-specific agents by 2026 [Gartner 2025], a figure we cite as directional only — but the governance substrate to answer "who is responsible?" for those agents is conspicuously immature. Importantly, in the system we describe, agent creation is not frictionless: binding requires a deliberate per-agent linkage step — registering the agent's public key and accepting the linkage verifiable credential (VC) — though the underlying identity verification is performed once per person, not per agent (Section 3.2.1). We frame the motivation accordingly — agents are *cheap to create but should be hard to attribute incorrectly* — and present the linkage step as the intentional counterweight to easy creation rather than an oversight. Sub-agents that share a parent agent's MCP (Model Context Protocol) server inherit the linkage without a separate step (Section 3.2.1).

The other half of the motivation is that agent operators have legitimate privacy interests. An individual or a small enterprise that runs an agent has a real stake in not exposing its legal identity to every platform the agent touches and every counterparty it transacts with. This is not a hypothetical worry. The platform economy has repeatedly demonstrated how identity, once disclosed to an intermediary, becomes an instrument of asymmetric power and commercial exploitation over the disclosing party [Zuboff 2019; Calo and Rosenblat 2017]. In an agent economy where agents act on behalf of many kinds of principals, the demand that identity be continuously visible to platforms and counterparties is precisely the demand whose costs the privacy literature has documented.

## 1.2 The thesis

Our central argument is a separation claim. *Association* — binding an agent to a verified legal principal — need not entail *disclosure* — revealing that principal to any business-layer participant. The mechanism that realizes the separation is *split-knowledge binding*: the identity half and the linkage half are held by two distinct government agencies, and no party in the business layer holds either. An agent can be bound yet anonymous to everyone except an authorized tracing authority acting under due legal process. The accountability that matters for legally consequential action survives; the identifiability that the dominant design treats as its precondition does not.

## 1.3 What we built

We present China's national identity layer for AI agents. Its core is real-identity association via anonymous authentication, in which an agent is bound to a verified principal without the principal's identity being disclosed to any business-layer party, accessed through a single embedded connector. Each agent maintains a local key pair that anchors the binding and also serves the agent's routine security functions — signing its actions and encrypting and decrypting its communications. Each agent additionally receives a nationally unified, human-memorable addressing identifier issued by the China Internet Network Information Center (CNNIC). The system is state-operated and built, with public launch scheduled for Q3 2026: the association and addressing services are operated by a joint venture majority-owned by the State Information Center (SIC) in conjunction with CNNIC, and the identity-binding layer composes with the National Network Identity Authentication Public Service, operated by the Ministry of Public Security (MPS).

## 1.4 The caveat, stated up front

We state the strongest objection before making any claim that might invite it. This is escrowed, conditional anonymity. It rests on trust in a state-held tracing channel, not on unconditional cryptographic anonymity. We do not present it as the latter, and a substantial part of this paper's purpose is to characterize that trade-off rigorously rather than to oversell it. A reader who finishes Section 4 should be able to state, more sharply than a critic could, exactly what the design does not protect against.

## 1.5 Contributions

We make five contributions, each defined in Section 2 and instantiated against the national system in Sections 3–4.

1. **Split-knowledge binding** — a coined mechanism for escrowed accountability in which the agent→principal mapping is divided into two components held by separate parties, such that neither party alone can re-identify, and the separation is institutional rather than cryptographic (Section 2.2).

2. **The ex-post attribution thesis** — the argued claim that for AI agent actions with legal consequences, only attribution-based accountability carries legal force, and that this has specific architectural requirements: a two-link chain (action→agent, agent→principal) composed by a legal proceeding (Section 2.1).

3. **The accountability surface** — a design concept capturing which agent actions leave identity-bearing traces: a tunable parameter ranging from no attribution through per-session, per-call, to per-message granularity, with a mapped trade-off between attribution strength and linkability (Section 2.3).

4. **A proportionality framework for identity escrow** — a decision structure grounded in legal proportionality that selects among three trust architectures (user-sovereign, state-escrow, cryptographic-threshold) via a decisive question (whose dominant threat is greater — private-sector or state?), two modifiers, and a boundary condition, and produces a falsifiable cross-border prediction (Section 2.4).

5. **The reflexive jurisdiction method** — an evaluative standard for identity-escrow architectures: apply the framework's rule-of-law condition to the deployment jurisdiction, state what the architecture determines (the *cost* of re-identification) and what the jurisdiction determines (what it *takes*), and identify the legal question the architecture cannot answer. The method is administered to the paper's own deployment in Section 4.4 (Section 2.5).

We are equally explicit about our non-goals. We do not claim unconditional anonymity, and we do not claim cryptographic novelty: the building blocks are mature, and our contribution is their applied composition and the governance framing around them. Nor do we argue that other jurisdictions should adopt this model.

---

## 2. Conceptual Framework

This section defines the paper's five conceptual contributions in abstract terms, before the system instantiation. Each is stated as a self-contained definition so that it can be cited independently of the Chinese deployment; Section 3 then describes how one national system instantiates them.

### 2.1 The ex-post attribution thesis

Four conceptions of agent accountability are in play in current research and practice, each distinguished by the mechanism on which accountability rests:

| Conception | Core mechanism | Sufficient for legal remedy? |
|---|---|---|
| (i) **Transparency-based** | Continuous observability of agent actions by operators or the public [Chan et al. 2024] | No — visibility without attribution does not assign legal responsibility |
| (ii) **Reputation-based** | Public, persistent identity and track record for counterparty trust decisions [Birch and Hoffart 2025] | No — reputation deters and informs but does not redress a harmed party |
| (iii) **Prevention-based** | Constraints, sandboxing, and authorization that block harmful action before it occurs | No — no preventive system catches everything; the residue is the set of harms requiring after-the-fact attribution |
| (iv) **Attribution-based** | A reliable, due-process-gated path from a harmful action to a definite legal principal | **Yes** — only conception (iv) carries legal force |

Our thesis is that for AI agent actions with legal consequences — fraud, tortious harm, contractual breach — conceptions (i)–(iii) are each valuable and complementary, but none is individually sufficient. Visibility without attribution does not assign legal responsibility; one can watch an anonymous actor in perfect detail and still have no defendant. Reputation deters and informs but does not redress: a damaged reputation is not a remedy to the party harmed. Ex-ante constraints fail open, because no preventive system catches everything, and the residue is exactly the set of harms that require after-the-fact attribution. Each of (i)–(iii) remains valuable, and each is *complementary* to (iv) rather than displaced by it. But only (iv) carries *legal force*, and legal force is precisely what an agent economy with real-world harms requires and what the other three cannot supply on their own.

We define **attribution-based accountability, for this domain, as the existence of a reliable, due-process-gated path from an agent action to a definite legal principal.** Three qualifiers carry weight. "Reliable" means the path works when lawfully invoked. "Due-process-gated" means it is not available to anyone at any time. "Definite" means it resolves to a specific principal rather than a probabilistic

match — though *definite* is not the same as *responsible*: a straw principal can be a definite legal person who is not the party actually responsible (Section 4.3.4).

Because the definition requires a path from an *action*, the architecture must supply two independently verifiable links: **action → agent** (per-action verifiable presentations, Section 3.2.4) and **agent → principal** (split-knowledge trace, Section 4.2). A legal proceeding composes the two. This thesis is domain-scoped, not universal: we claim it for legally consequential agent actions and explicitly do not claim that conceptions (i)–(iii) should be displaced where they are the dominant need.

We ground these conceptions in the accountability literature [Bovens 2007; Nissenbaum 1996; Wieringa 2020] while noting that the existing frameworks — Bovens' forum–actor–phases structure, Nissenbaum's answerability conception, Wieringa's systematic survey of algorithmic-accountability definitions — were developed for human and organizational accountability rather than agent accountability. The four-conception table above is a domain-specific mapping, not a replacement for those frameworks.

## 2.2 Split-knowledge binding

Split-knowledge binding is a mechanism for escrowed accountability in which the agent→principal mapping is divided into two components held by separate parties, such that (i) neither party alone can map an agent to a person, (ii) re-identification requires both components and is gated by a defined legal process, and (iii) the separation is institutional rather than cryptographic — a governance barrier between organizations, not a mathematical barrier between keys.

The term adapts "split knowledge" from cryptographic key custody [NIST SP 800-57], where it denotes dividing a secret into components so that no single party holds the whole and only collaboration reconstructs it. We transpose the principle from secrets to the identity-binding itself: the two halves are *knowledge of the verified identity* and *knowledge of the agent-to-pseudonym linkage*, held by different agencies. "Split-knowledge binding," in this identity-binding sense, is introduced by this paper.

The key departure from prior cryptographic constructions is that the split operates at the institutional layer. Where identity-escrow schemes divide capability across cryptographic escrow agents [Kilian and Petrank 1998], split-knowledge binding divides knowledge across real agencies whose separation is maintained by organizational mandate, administrative independence, and distinct oversight structures. The limit — that a state compelling both agencies defeats the split — is the same limit that cryptographic threshold escrow faces when all share-holders fall within one jurisdiction. The difference is that an institutional split makes the compulsion *visible*: it leaves a procedural record across two agencies where a purely cryptographic split may not. This institutional transposition — not a new cryptographic primitive — is the mechanism's contribution.

The architecture of Section 3 instantiates this mechanism with two specific Chinese agencies: MPS holds identity without the agent linkage, and SIC holds the linkage without the identity. The user is the only party holding both halves during the binding ceremony; afterward, neither agency alone can re-identify. The pairwise reference at the center of the construction is an instance of the sector-specific pseudonym familiar from eIDAS [Regulation (EU) 2024/1183], cited to show the choice is principled rather than ad hoc.

The mechanism's property — escrowed (conditional) anonymity — is the unlinkability of an agent's identifier to its legal principal for all business-layer participants and for either government agency acting alone. Re-identification requires combining the two halves, which only a legal authority can do by separately compelling both agencies. The pairwise reference affords no cross-sector correlation: SIC cannot use it to track the person across other services, and other services receive different references for the same person.

The property is *escrowed* because the tracing capability exists, and *conditional* because the protection is structural, not cryptographic: there is no mathematical control that refuses an authorized-but-improper recombination once both agencies are compelled. We state this in the same breath as the property, because the difference between a claim that survives scrutiny and one that does not lies precisely here.

This construction sits within an established cryptographic lineage — identity escrow [Kilian and Petrank 1998], key-escrow taxonomies [Denning and Branstad 1996], selective traceability [von Ahn et al. 2006], revocable anonymity [Köpsell et al. 2006], anonymous credentials with optional revocation [Camenisch and Lysyanskaya 2001], mutual accountability [Daza et al. 2022], and non-frameability in anonymous systems [Backes et al. 2014] — to which we return in Section 6.4. We introduce no new cryptographic primitive. Our contribution is an applied, national-scale, agent-oriented realization within this well-studied family, inheriting both its core mechanism and its central critique. We also inherit the systemic-risk framing of the exceptional-access debate [Abelson et al. 1997; Abelson et al. 2015] and answer it in Section 4.1: the split-knowledge design answers the concentrated-target argument, the institutional-rather-than-protocol mechanism partially blunts the complexity argument, and the precedent argument is faced, not answered, in Section 4.4.

## 2.3 The accountability surface

The accountability surface is the set of agent actions for which an identity system produces a verified, retained record binding the action to a bound agent identity. It is a design parameter, not a fixed property.

The surface can be placed at different points on a granularity spectrum: no attribution (no actions leave identity-bearing traces), per-session (a session-key presentation at connection time), per-call (a verifiable presentation generated and verified for each accountability-requiring invocation), or per-message (each in-session message is additionally signed). Finer surfaces strengthen attribution — more actions are traceable to an agent — but increase linkability: within any one service, the same agent identifier links all of that agent's VP-carrying calls over time. Coarser surfaces preserve more privacy and leave more room for non-attributable action, but create attribution gaps. The design question is *where* to place the surface, not *whether* accountability exists.

Accountability-surface placement also determines when credential revocation takes effect. If the surface is at the call level and VPs are generated and verified per call, revocation takes effect at the next VP-verified call — the mechanism that makes revocation real rather than notional.

In the system described in this paper, the surface is placed at the per-call level for service gateways that require accountability; gateways that do not require accountability may accept calls without a VP (Sections 3.2.3 and 3.2.4). This is a design choice, not an architectural necessity: a different deployment could place the surface elsewhere.

## 2.4 A proportionality framework for identity escrow

Three trust architectures — user-sovereign, state-escrow, and cryptographic-threshold — are competing answers to a single proportionality question: given a legitimate accountability aim, which architecture achieves it with the least intrusion on the relevant party at risk?

The question is a *proportionality* question of the kind legal systems already adjudicate — proper purpose, suitability, necessity (least-restrictive-means), and balancing [Barak 2012; Alexy 2002]. We frame the three models as points in a structured design space.

**The user-sovereign model** (eIDAS 2.0 / EUDI Wallet). The user holds their own identity credentials and exercises selective disclosure, deciding per interaction what identity attributes to reveal [Regulation (EU) 2024/1183]. Its strengths are user autonomy, resistance to state over-reach, and alignment with data-minimization norms. Its limitations for *agent* accountability are threefold: the user can choose to disclose nothing, which makes accountability voluntary; per-transaction selective disclosure may be impractical for autonomous agents making thousands of decisions; and there is no built-in mechanism for lawful access when the user is unwilling or unavailable.

**The state-escrow model** (this work; instantiated via split-knowledge binding, Section 2.2). Disclosure is withheld from all business-layer participants and escrowed to the state, released only through due process. Its strengths are that accountability does not depend on user cooperation, and that privacy is structural — nobody in the business layer can see identity — rather than discretionary, where the user chooses each time. Its limitations are that re-identification remains possible to a state that compels both agencies; that it carries function-creep risk; and that it does not travel across jurisdictions.

**The cryptographic-threshold model.** De-anonymization requires the cryptographic cooperation of $k$ of $n$ parties; an improper trace is computationally impossible without a quorum [Köpsell et al. 2006; Camenisch and Lysyanskaya 2001; Daza et al. 2022]. Its strength is that abuse-resistance is enforced by mathematics. Its costs are national-scale multi-party key management, rotation, and recovery across institutions; quorum availability as an operational failure point for legitimate traces; and the fact that the quorum parties would in practice still be state-appointed, so the quorum's independence is itself institutional, not mathematical. On abuse-resistance alone, the threshold model dominates, and we do not pretend otherwise.

**Decision structure.** The framework selects a default via a decisive question — whose dominant threat is greater? — and modifies it by two further dimensions. When the principal's dominant risk is *private-sector* exploitation (platforms, counterparties, surveillance capitalism), escrow's structural privacy dominates, because it removes identity from exactly the parties that would exploit it. When the principal's dominant risk is *the state*, user-sovereignty dominates, because escrow concentrates exactly the capability the principal most fears. Two modifiers can flip the default: the legal-consequence weight of the domain (supplied by the ex-post attribution thesis, Section 2.1), and the rule-of-law quality of the jurisdiction. A strong private-sector-threat case is overridden if judicial oversight cannot be shown to constrain the escrow authority — in which case neither model is safe, and the honest answer of the framework is "do not deploy escrow here."

**Prediction.** The framework makes a specific, falsifiable prediction. Consider a mixed cross-border scenario in which Jurisdiction X has strong rule-of-law and Jurisdiction Y does not, and an agent bound in X interacts with an agent bound in Y. The framework predicts that escrow is defensible intra-X; that X must not honor Y's tracing requests absent X-level due process; and that, as a consequence, the agent economy fragments at the rule-of-law boundary.

The framework classifies; it does not compute. It cannot deliver a verdict in a genuinely balanced case. What it does is make the grounds of disagreement explicit — and that is the contribution.

## 2.5 The reflexive jurisdiction method

The reflexive jurisdiction method is an evaluative standard for identity-escrow architectures: apply the proportionality framework's rule-of-law condition to the deployment jurisdiction, state what the architecture determines and what the jurisdiction determines, and identify the legal question the architecture cannot answer.

The method separates two questions that are often conflated. The architecture determines what re-identification *costs*: how many parties must be compelled, what procedural record is created, and what technical barriers exist to improper tracing. The jurisdiction determines what re-identification *takes*: whether the mechanisms of judicial oversight credibly constrain the tracing authority, and whether the dominant threat in that jurisdiction is private-sector or state. The former is an architectural fact; the latter is an institutional fact no architecture can supply.

The method does not produce a verdict. It produces a structured question: the jurisdiction-specific legal analysis that an actual appropriateness judgment would require. Its value is that it makes the grounds of disagreement explicit and sets a standard other identity-escrow papers can adopt — administer your own framework to your own deployment. We apply the method to this paper's deployment in Section 4.4.

The five contributions form a chain. The ex-post attribution thesis (C2) identifies which agent actions require accountability and why only attribution-based mechanisms carry legal force. Split-knowledge binding (C1) supplies the mechanism that delivers that attribution without business-layer disclosure. The accountability surface (C3) specifies *where* the mechanism leaves identity-bearing traces — which actions are covered and which are not. The proportionality framework (C4) selects among the resulting trust architectures and identifies the conditions under which escrow is and is not defensible. The reflexive jurisdiction method (C5) converts those conditions into an evaluable standard and applies it to the paper's own deployment, surfacing the jurisdiction-specific question that no architecture can answer for itself.

The remainder of the paper instantiates this framework against a specific national system. Section 3 describes the architecture and threat model; Section 4 applies the framework analytically.

---

# 3. A National Instantiation

## 3.1 Institutional context

Because the system's governance claims rest on the institutional character of the participating agencies, we state their roles and authority explicitly before the architecture.

**Actors and setting.** We use the following roles throughout.

- **Agents** are semi-autonomous software entities that act on behalf of legal principals.
- **Legal principals** are natural persons or legal entities that bear legal responsibility for an agent's actions.
- **The agent runtime** is owned and operated by the principal: agents are locally deployed — on a PC, a phone, or the principal's own cloud — with the embedded connector installed inside this owner-controlled environment. There is no central platform hosting agent execution.
- **Third-party applications and service gateways** are external software that invokes agent capabilities or that agents call; service gateways are the verification points at which accountability-requiring services check an agent's credentials (Section 3.2.4).
- **The addressing registry** issues and resolves the agents' nationally unified identifiers (operated by CNNIC).
- **Judicial and law-enforcement authorities** are the only parties authorized to trace an agent to its principal.
- **The three participating agencies** are (i) the National Network Identity Authentication Public Service ("National Identity Service," operated by the Ministry of Public Security, MPS), which holds the mapping from a pairwise reference to a true identity but never sees the agent or its linkage; (ii) the State Information Center (SIC), which holds the mapping from that reference to the agent's public key and identifier but never sees the true identity; and (iii) the China Internet Network Information Center (CNNIC), which operates the addressing registry but holds no binding data at all. Re-identification requires both identity and linkage halves, recombined only by a legal authority that separately compels MPS and SIC.

**The State Information Center (SIC)** is a public institution directly affiliated with the National Development and Reform Commission (NDRC), China's central macroeconomic planning agency under the State Council. Established in 1987, SIC operates the Administration Center of China E-government Network. In the system described here, the agent-identity infrastructure — the linkage registry and the connector — is operated by a joint venture majority-owned by SIC, with SIC retaining institutional authority and holding the agent-linkage half of the split-knowledge binding. Throughout this paper, "SIC" refers to this institutional umbrella unless the operating entity is distinguished.

The **China Internet Network Information Center (CNNIC)** is the national Internet registry, established in 1997 and operating under the Ministry of Industry and Information Technology (MIIT). CNNIC administers the .cn country-code top-level domain and China's fundamental Internet resources. In this system, CNNIC transposes its existing registration authority to the agent-addressing layer: it issues and resolves human-memorable agent identifiers.

The **National Network Identity Authentication Public Service**, operated by the **Ministry of Public Security (MPS)**, is China's official online identity authentication infrastructure. Codified by the Measures for the Administration of the National Network Identity Authentication Public Service (Order No. 173, effective July 2025), it provides privacy-by-design anonymous authentication: in its anonymous mode — the mode used by the integration described here — a relying party authenticating a user receives only a validity confirmation and a sector-specific pairwise reference rather than the user's plaintext identity; plaintext identity can be returned only to specifically authorized government agencies under law. In this system, the MPS service is the identity half of the split-knowledge binding — it verifies the principal and holds identity, but never sees the agent or its linkage. SIC integrates as an authorized relying party of this national infrastructure for this purpose; the split-knowledge architecture is the deployed integration contract of the MPS service's anonymous mode, not an escrow design improvised for this system (Section 4.2.3).

These three agencies are organizationally and administratively distinct: they report to different commissions (NDRC, MIIT, and the State Council respectively), operate under different legal mandates, and are subject to different oversight structures. The split-knowledge binding derives its institutional strength from this separation of powers, and its limits from the fact that all three are arms of one state — a boundary we analyze in Sections 4.2.3 and 4.3.1.

## 3.2 Architecture

This section describes how the national system instantiates the conceptual framework of Section 2. Readers seeking technical substance will read it most closely, and the binding ceremony — where the split-knowledge binding is recorded, who holds it, who can resolve it, and under what authorization — is its center. We are forthright that the cryptographic components are off-the-shelf: the novelty is in the composition and its governance properties, not in the parts.

### *3.2.1 Overview: identity association through a single connector*

The system's core is a single function — real-identity association with anonymous authentication (Section 3.2) — reached through a single embedded connector: an MCP server provided by SIC and installed in the agent's own runtime environment (Section 3.2.3). The association is exercised through credentials: the linkage VC issued at binding (Section 3.2.1) is used to generate a VP for each accountability-relevant action, which is how an agent proves, action by action, that it is bound to a verified principal without revealing who (Sections 3.2.3 and 3.2.4). Two supporting facts complete the picture, and we state them without ceremony. First, each agent maintains a local key pair; beyond anchoring the linkage credential, the same key material can sign the agent's other actions and encrypt and decrypt its communications — an add-on to the credential mechanism, not the system's main action. Second, each agent receives a nationally unified, human-memorable identifier of the email-like form `agent-name@registry-suffix`, issued and resolved by the China Internet Network Information Center (CNNIC), mapping to the agent's current endpoint and public key. Figure 1 shows the split-knowledge binding and trace that constitute the system's privacy-defining mechanism.

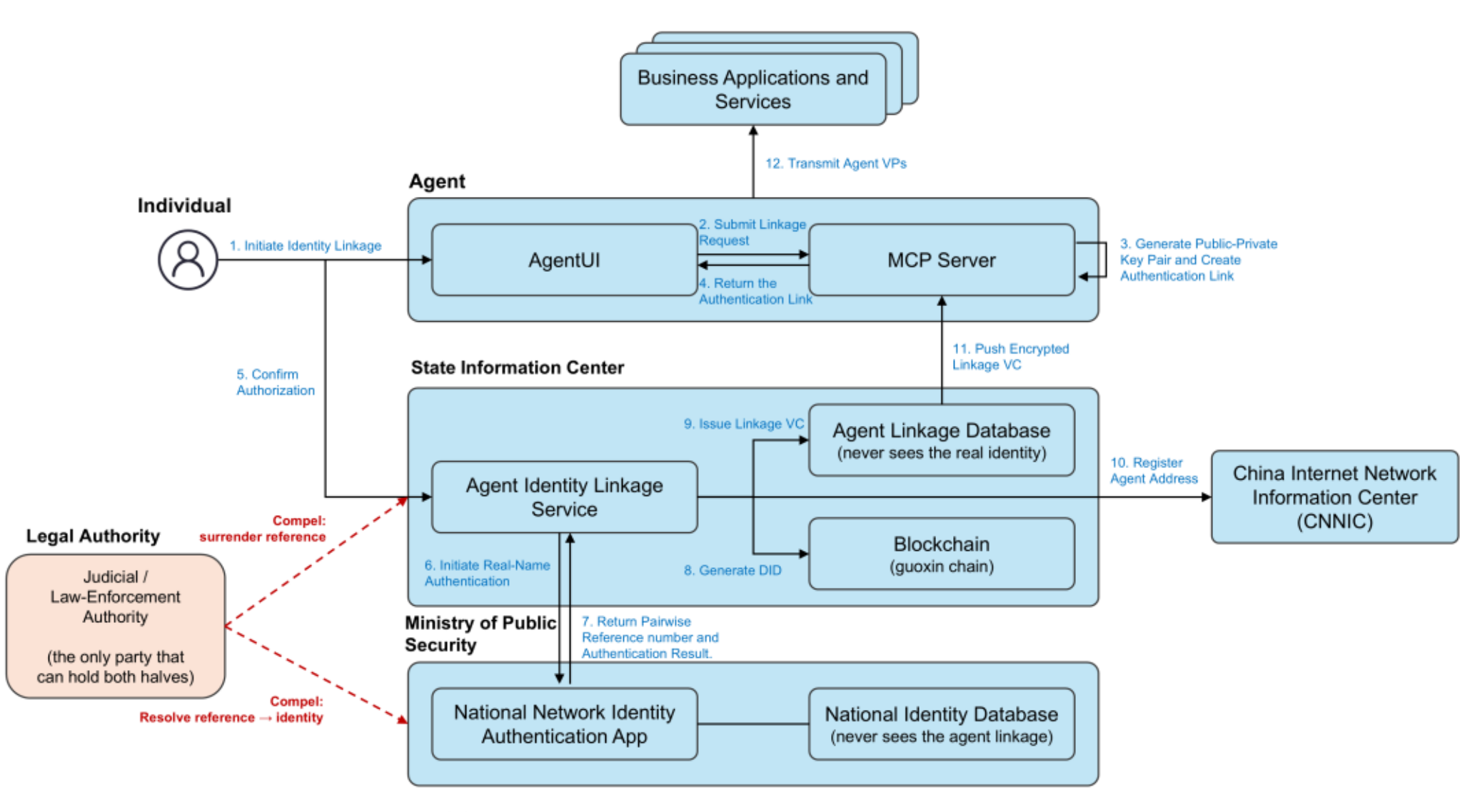

*Figure 1. Split-knowledge binding and trace.*

**Figure 1. Split-knowledge binding and trace.** An individual authorizes an agent's identity linkage through SIC's Agent Identity Linkage Service, which redirects to the Ministry of Public Security's National Network Identity Authentication App for real-name verification and receives back only a pairwise reference and validity result — never the true identity. SIC's Agent Linkage Database issues the linkage VC and registers the agent's address with CNNIC; the VC is pushed to the agent's MCP server, which presents verifiable presentations (VPs) to business applications and services per call. SIC holds linkage-without-identity; MPS holds identity-without-linkage; the pairwise reference (disclosed to no one, in no VC) is the only join key. The dashed arrows show the trace path: a legal authority separately compels SIC to surrender the reference and MPS to resolve it to an identity — the two disclosures meet only inside the legal authority, never inside the system. Dashed arrows are policy-and-law-gated disclosures, not technically enforced controls, and the figure depicts no combinable end-to-end audit.

### *3.2.2 Real-identity association with anonymous authentication*

#### The binding ceremony

Binding proceeds in two phases. **Phase 1 — one-time identity verification and DID issuance.** The person, through SIC's agent-identity application, is redirected — opaquely — to the MPS National Identity Service, which performs identity verification (including liveness and facial checks) and returns a **pairwise reference number.** This reference is deterministic and stable for the pairing of *this person* with *SIC*, but different for the same person toward any other relying party. SIC receives the reference together with a confirmation of validity and issues a DID to the person without learning the true identity. This verification is performed **once per person**, not per agent.

**Organizational principals.** Phase 1 as just described is the natural-person path: individual principals are verified through the MPS National Identity Service. When the principal is a legal entity — an enterprise or other registered organization — identity verification is performed instead against the **China Organization Data Service (CODS)** of the State Administration for Market Regulation (SAMR), which manages enterprise and organization registration data in China, and which occupies for organizations the identity-authority role that the MPS service occupies for individuals. The rest of this paper describes the individual path, for two reasons: it keeps the exposition to one verification authority, and it is the privacy-critical case — organizational identity is largely a matter of public record, while personal identity is not. References to the MPS service should be read with this scoping in mind.

**Phase 2 — agent linkage via MCP server.** The person installs the SIC-provided MCP server — the embedded connector (Section 3.2.3) — in their own runtime environment and links it to their DID; the MCP server is part of the solution, and each locally deployed agent operates through it. Each agent — not the DID — maintains a **local key pair** (a public key and a private key). The agent's public key is registered with SIC, which issues a verifiable credential (VC) binding the person's DID to the agent's public key. The private key never leaves the agent and is used to sign actions (Section 3.2.4) and to decrypt traffic encrypted to the agent's public key. Because identity verification occurs only in Phase 1, linking a new agent does **not** require re-verification: the person uses their already-issued DID, and SIC issues a new linkage VC for the new agent's public key. The per-agent ceremony is therefore a **per-agent linkage step** — registration of a new public key and issuance of a new VC — rather than a re-run of identity verification.

The binding is split-knowledge: MPS holds `pairwise-reference → true identity`; SIC holds `pairwise-reference → person-DID → agent-public-key → agent-identifier`. The pairwise reference is the only join key, held solely by MPS and SIC, and present in neither the DID document nor any VC. The agent receives a VC attesting that it is "bound to a verified principal" without containing the principal's identity or the pairwise reference; the agent, or the MCP server acting on its behalf, uses this VC to generate verifiable presentations (VPs) that prove linkage to service gateways (Section 3.2.3).

**Sub-agent inheritance.** Agents that share the same MCP server are linked to the same person's DID. Sub-agents — agents spawned under a parent agent — that share the parent's MCP server inherit the linkage and the parent's key pair, and can invoke VP creation to demonstrate their bound status to service gateways, without requiring a separate linkage step or a key pair of their own. A direct consequence follows: signed actions and VPs from a sub-agent are cryptographically indistinguishable from the parent's, so attribution resolves to the shared-key cluster — the parent and all its sub-agents together — not to the specific sub-agent that acted. We treat this as a limitation, not a feature (Section 7, L10).

**Honest consequences.** While identity re-verification is avoided for new agents, the binding friction that remains is that each new agent requires a deliberate linkage action by the principal (registering the agent's public key with SIC). We frame this as an intentional accountability feature rather than an oversight: agents are *cheap to create but should be hard to attribute incorrectly.* A person who never links an agent is not responsible for it; the linkage step is the point at which accountability is knowingly accepted. Notably, the one-time nature of identity verification weakens the anti-straw-principal argument relative to a full per-agent verification regime — a cost we analyze honestly in Section 4.3.4. Finally, because the pairwise reference is stable per (person ↔ SIC), all of a principal's agents share it, making them linkable by SIC into a pseudonymous profile (Sections 2.2 and 4.3.5).

**Voluntariness.** Binding is voluntary under present law. No legislation currently requires an agent to be bound to a principal through this system, and operating an unbound agent is not, by this system, prohibited; what binding provides is a credential that accountability-requiring service gateways can demand (Section 3.2.3), so the practical pressure to bind scales with how much of the service economy requires verification. We state this plainly because voluntariness is not incidental to the fairness analysis: a voluntary credential and compulsory registration are categorically different objects, and Sections 4 and 7 should be read under the voluntary regime that presently obtains. Whether future legislation will mandate binding is beyond the authors' knowledge or control; if it does, the appropriateness analysis of Section 4.4.1 must be re-administered under that changed premise.

### The decoupling mechanism (the technical heart)

We specify the four things a reader needs to evaluate the privacy claim: who stores what, how binding proceeds, how a trace proceeds, and what cryptographic methods are used.

**Storage — who holds what.** There are two separated stores. The MPS National Identity Service retains a mapping `pairwise-reference → true identity`; it is its half, and it never sees the agent or the person's DID. SIC retains `pairwise-reference → person-DID → agent-public-key → agent-identifier`, materialized by the linkage VC it issued; it is its half, and it never sees the true identity. No store contains both identity and agent.

**Binding protocol.** User-driven and two-application: SIC application → (opaque redirect) → MPS service verifies (facial and other) → returns the pairwise reference → SIC records the linkage and issues the linkage VC (binding the person's DID to the agent's public key). The pairwise reference is the join key, held only by MPS and SIC, present in neither the DID document nor any VC.

**Resolution (trace) protocol.** Strictly sequential and authority-mediated (detailed in Section 4.2): a legal authority compels SIC, which surrenders the pairwise reference for the named agent and then holds no further knowledge; the legal authority then compels the MPS National Identity Service, which resolves the reference to an identity. **The legal authority is the only party that ever holds both halves; neither MPS nor SIC does, even during a trace.**

**Cryptographic binding method.** A person-held DID, the agent's public key, and a W3C verifiable credential binding the DID to the agent's public key [W3C VC 2.0]; and a pairwise pseudonymous reference — an eIDAS-style sector-specific identifier [Regulation (EU) 2024/1183] — as the withheld join key. The anonymous-credential lineage is that of Camenisch and Lysyanskaya [2001]; the pairwise-pseudonym construction is the eIDAS one.

It follows that no business-layer participant — applications, service gateways, addressing network, or other agents — can resolve the linkage, and that neither government agency can resolve it alone; only a legal authority combining two separately compelled disclosures can. The per-relying-party uniqueness of the pairwise reference further prevents cross-sector correlation even by SIC. Section 5 states the system's launch status; all mechanisms described are implemented and operational under internal testing.

### Addressing and the "real-name, not real-info" principle

The authentication system knows *that* a principal is real, not *which* principal: during the anonymous-authentication flow it asserts validity without transmitting identity attributes to the agent-identity infrastructure. This is distinct from pseudonymity. The agent is not linked to a persistent pseudonym that could be correlated across contexts; its public identifier is the CNNIC-issued address, which reveals nothing about the principal. The "real-name, not real-info" mode is not a convention of this system alone — it is the privacy-by-design behavior codified for the MPS National Network Identity Authentication Public Service, under which platforms authenticating a user through the service may not demand the user's plaintext identity [Measures for the Administration of the National Network Identity Authentication Public Service 2025]. We return to the significance of this in Section 4.2.3.

### *3.2.3 Access model: the single connector*

All agent access to the identity layer is mediated through a single embedded connector: the MCP server provided by SIC and installed within the principal's own runtime environment (Section 3.2.1). Third-party applications invoke the connector for credential association (VC/VP), cryptographic signing and encryption, and addressing lookup. The connector is the architectural enforcement point: it ensures that no agent can bypass the identity layer.

Three implemented properties of the connector matter for the rest of the paper. First, agent calls to service gateways that require accountability carry a **real-time-generated VP with selective disclosure**; the gateway verifies the call as originating from a bound, identity-linked agent, without learning the identity and seeing only the attributes the VP exposes. Gateways that do not require accountability may accept calls without a VP. Second, gateways **retain the verified VP records** for future audit or legal action — this is the implemented evidence mechanism with which Section 3.2.4's accountability chain composes: a retained VP record (call → agent) plus a split-knowledge trace (agent → principal). Third, because VPs are generated and verified per call, **credential revocation takes effect at the next VP-verified call** — the mechanism that makes the reference-level revocation of Section 4.3.6 real rather than notional.

Three honest consequences follow. Attribution granularity is the call or session, via the VP record; whether in-session message content is additionally signed is a deployment detail (Section 3.2.4). The retained records form a **distributed, subpoenable metadata corpus by design**: each service holds an audit trail of which pseudonymous agents called it and when. This is the audit feature working as intended, and we name it here and in Sections 2.2 and 4.3 rather than let a reader discover it. Finally, within any one service the same agent identifier links that agent's VP-carrying calls over time — an inherent property of verifiable per-call identity-linkage; selective disclosure limits *attribute* leakage, not call-level linkability.

### *3.2.4 Action attribution: per-action VPs and signed actions*

Section 2.1 defines accountability as a path from an *action*, but Sections 3.2.1–3.2.3 architect only the *agent* → *principal* link. The first link — *action* → *agent* — must be made explicit, or the definition outruns the system. The per-call VP mechanism of Section 3.2.3 already supplies it: each accountability-requiring call leaves a verified, retained VP record that attributes the call to a bound agent. Action attribution adds one further mechanism: agent actions and messages can also be **cryptographically signed** with the agent's private key, whose corresponding public key the SIC-issued VC anchors to the person's DID (Section 3.2.1), so that a counterparty holding a signed interaction record possesses non-repudiable evidence that *this agent* performed *this action*. The accountability chain thus has two independently verifiable links — **a per-action VP or signature (action → agent)** and **the split-knowledge trace via the linkage VC (agent → person-DID → principal)** — that a legal proceeding composes.

We bound the claim honestly. The system guarantees attributability of VP-carrying and signed interactions; off-protocol harms — an agent's side effects in the world that pass through neither mechanism — attach to the agent only through conventional evidence, and we say so. Signing also reinforces non-frameability (Section 4.2.5): an innocent principal cannot be attributed an action whose signature their agent's key never produced. Whether in-session message *content* is additionally signed (which would move attribution granularity from the call/session level to the message level) is a deployment detail to be confirmed in the post-launch revision, which will also report verification volumes where permissible.

---

## 3.3 Threat model

### *3.3.1 Adversaries*

We consider the following adversaries, each distinct.

- **A1 — A curious or malicious service operator** — an application or service-gateway operator — seeking to learn which principal is behind which agent. Because agents are locally deployed and the runtime belongs to the principal (Section 3.1), the parties in a position to observe an agent are the services it calls, not a hosting platform.
- **A2 — Colluding third-party applications** attempting correlation-based de-anonymization across an agent's interactions (analyzed in Section 4.3.7).
- **A3 — A passive network observer** attempting to link agent identifiers to principals through traffic patterns. (We list this adversary for completeness and then, consistent with Section 3.3.3, hand it off as out of scope: network-layer traffic analysis is a separate and well-studied problem for which we assume standard mitigations.)
- **A4 — A malicious agent operator** seeking to evade accountability for harmful agent actions.
- **A5 — A straw-principal launderer.** An instigator who binds agents through a paid, coerced, or identity-theft-recruited *genuine* principal, so that lawful tracing resolves — correctly, by the system's own lights — to the wrong or judgment-proof person. Non-frameability (Section 4.2.5) does not address this adversary, because the binding is cryptographically genuine; the deception is social. Every real-name registration regime has spawned a black market in verified accounts [Lee and Liu 2016; Thomas et al. 2013], and we assume this one will too. We analyze it in Section 4.3.4.
- **A6 — An over-reaching or compromised tracing authority.** This is the central adversary, and the one most work in this space underweights. It may attempt to de-anonymize without due process, to conduct bulk surveillance, or to leak association data. Its formal analogue is the “malicious manager” of the Mutual Accountability Layer [Daza et al. 2022]; its real-world pattern is the expansion of surveillance mandates documented in the function-creep literature [Koops 2021; Solove 2011].

- **A7 — A third party harmed by an agent, seeking recourse.** This is not an adversary in the security sense but the actor whose interests a fairness audience centers, and we include it in the threat model deliberately. The architecture makes *private* de-anonymization impossible by design, so a victim of an anonymous agent — a defrauded counterparty, a harmed bystander — has no path to redress except by triggering a lawful tracing process; they cannot themselves discover who is responsible. We do not leave this silent, because at a fairness venue omitting the harmed party reads as never having considered victims. We analyze it in Section 4.1 as a deliberate, defended design choice with an honestly stated cost.

### 3.3.2 Trust assumptions

We rely on the following assumptions, stated so they can be challenged.

- **The MPS National Identity Service correctly verifies legal identity** at binding time and returns only a pairwise reference; no identity attributes flow to SIC.
- **SIC never receives the true identity and never discloses the pairwise reference** — not in the DID document, not in any VC, not to any party.
- **The two agencies do not collude off-process** to recombine their halves outside lawful process. This is a *governance and legal* trust assumption, not a technical guarantee — the architecture cannot prevent a state that compels both. It is the model’s load-bearing assumption, and Sections 4.3.1 and 4.4 address it directly rather than hiding it.
- **The principal’s own device and the applications’ integrity at binding time are not compromised.** We do not leave this to the standard endpoint assumption alone. Implemented precautions include mechanisms to prevent host hijacking during binding, to ensure the agent’s public key registered with SIC is genuinely the agent’s own (preventing substitution that would link the principal to an agent they do not control), and to protect the VC transfer against interception. These raise the bar for endpoint compromise but do not eliminate it, and we do not claim otherwise.
- **The pairwise reference is generated so that it does not itself leak identity and is not reused across relying parties** (per-relying-party uniqueness).

### 3.3.3 Declared out of scope

Honesty about scope is part of the method. We declare out of scope: network-layer traffic analysis and metadata correlation (a separate, well-studied problem; we assume standard network-layer mitigations); compromise of the principal's own device or keys (endpoint security is orthogonal); and collusion between the tracing authority and a business party (the tracing authority is assumed to follow legal process — if it colludes, the model breaks, and we name this as a genuine limitation in Section 4.3.3 rather than pretending the architecture prevents it). These are the standard exclusions of the revocable-anonymity literature, and we adopt them as such rather than inventing convenient boundaries.

---

## 4. The Central Trade-off: Escrowed Accountability

This is the intellectual center of gravity of the paper, and the contribution a fairness-and-accountability audience will judge most carefully. We affirm first what the design buys (Section 4.0), then concede its limits (Sections 4.1–4.3), then compare it even-handedly against the user-sovereign and cryptographic-threshold alternatives (Section 4.4). We never claim the problem is solved.

### 4.0 What the design actually buys

Before any concession, the question every reader will otherwise answer uncharitably: after all the limits, what does a principal concretely get from split-knowledge that a single identity database with an access policy would not provide? Four things, each verifiable against Section 3.

First, **every business-layer party is structurally, not discretionarily, excluded.** The applications, the service gateways, the counterparties, and the addressing network each hold nothing that resolves to an identity — nothing *identity-bearing* to leak, to sell, or to be subpoenaed for. What they do hold is pseudonymous: the retained VP records of Section 3.2.3 are a subpoenable metadata trail, but one that yields agent-level pseudonyms, never the principal's identity — no business-layer subpoena, singly or in combination, can cross the split-knowledge boundary. This is a property of what they store, not of a policy they promise to follow.

Second, **each agency alone is defeated.** A curious insider, a data breach, or an improper single-agency demand yields, at MPS, identities with no agent linkage, and at SIC, linkages with no identity. A single-database design fails every one of these cases; this design fails none of them.

Third, **re-identification costs two separate legal compulsions against two separate agencies and leaves two independent paper trails.** This raises the procedural floor and creates evidence of systematic abuse even in the absence of a combinable audit (Section 4.2.2).

Fourth, **pairwise references block cross-sector correlation of the principal.** No business party and no other registry can join the person-level reference against other services' records. The *agent's* public identifier is, by contrast, globally stable

— a correlation channel at the agent level that we analyze and bound in Section 4.3.7.

We are disciplined about what this paragraph does not do: it does not reclaim anything that Sections 4.1–4.3 concede. It states the residual; the concessions then bound it.

## 4.1 This is conditional anonymity, and that is a design choice

We begin by stating a regulatory reality that the cryptographic-anonymity literature, for sound reasons internal to its project, does not take as given — but that any deployed national accountability infrastructure must. No major jurisdiction permits fully anonymous, untraceable *digital* financial or contractual activity at scale: cash aside, remote financial services operate everywhere under AML/KYC obligations, and anonymous bank accounts have been eliminated under the FATF baseline. No jurisdiction that has addressed the question permits fully anonymous agent behavior with legal consequence. Every regulatory framework that speaks to digital accountability — the EU Data Retention Directive's legacy, the US CLOUD Act, China's Cybersecurity Law, and the compliance baselines they represent — requires some form of lawful access. The design question for a real, deployed agent-accountability system is therefore not *whether* a tracing path exists but *how it is constrained*, and whether the constraints are structural rather than merely promissory. Escrowed anonymity via split-knowledge binding — under which tracing requires two separate legal compulsions against two separate agencies, and no business-layer party ever sees identity — is a strong point on the privacy frontier available under that compliance baseline, though not its maximum: Section 2.4 concedes that a cryptographic-threshold design would exceed it on abuse-resistance under the same baseline. The paper's contribution is not arguing against an impossible unconditional model but making the escrow *as constrained as the regulatory environment permits* and analyzing honestly what remains unconstrained.

### *4.1.1 The concession*

A tracing authority that can resolve identifier → principal means anonymity is conditional on trusting that authority. This is not a bug discovered late; it is the design premise. We place it within its known lineage rather than pretending it is new: key escrow (the Clipper-chip debates), revocable anonymity [Köpsell et al. 2006], group-signature openers, and selectively traceable anonymity [von Ahn et al. 2006]. That lineage carries a standing critique — single point of trust, potential for abuse — which we do not dispute [Daza et al. 2022].

The most important item in that critique, for an escrow paper, is the technical community's standing case against exceptional access [Abelson et al. 1997; Abelson et al. 2015]. We cite it because omitting it from an escrow paper would read as either ignorance or evasion, and we answer it in three parts, saying which is which. (i) The **concentrated-target** argument — that exceptional-access mechanisms create a single high-value point of attack — is genuinely answered by the two-agency split: there is no single recovery key and no single database to steal, and compromising one agency yields half of a binding that is useless without the other. This is a point in the design's favor, and we argue it as such. (ii) The **complexity** argument — that exceptional access adds dangerous system complexity — is partially blunted, because the mechanism here is institutional separation rather than a new cryptographic protocol surface; we do not claim it is fully answered. (iii) The **jurisdictional-precedent** argument — that building lawful-access infrastructure normalizes and entrenches it — is *not* answered by the architecture at all. It lands in Section 4.4.4, where we face it for our own deployment rather than deflecting it.

#### *4.1.2 The affirmative argument*

For an accountability infrastructure intended to carry legal force, *some* lawful tracing path is a requirement of the problem domain, not a defect of the solution. A system in which no one, ever, under any circumstance, can trace an agent to its principal does not provide accountability; it provides immunity. The real design question is therefore not *whether* tracing exists but *how* the tracing power is constrained — which is the subject of Section 4.2.

## 4.2 Constraining the tracing power

We describe the mechanisms that constrain the tracing authority; all are implemented and operational under internal testing.

#### *4.2.1 The trace sequence: two separately compelled, single-party-insufficient disclosures*

We correct at the outset any impression of a cryptographic threshold. The implemented mechanism is *not* a threshold or co-signature scheme. It is **two sequential, independently compelled disclosures** at two agencies, recombined only inside the legal authority. Calling it a threshold scheme would be an overstatement a reader could falsify against Section 3.2, and we do not make it.

A trace proceeds in two steps. First, a legal authority compels SIC, which surrenders the pairwise reference for the named agent and then holds no further knowledge. Second, the legal authority compels the MPS National Identity Service, which resolves the reference to an identity. SIC learns nothing after the first step. Neither MPS nor SIC ever holds both halves, even during a trace; only the legal authority does. The system itself never re-identifies anyone — it only enables a court to. The protection against abuse is structural and procedural: an adversary needs two separate legal compulsions against two separate agencies (MPS and SIC) and must itself be the authority that correlates the results, which raises cost and creates a two-step paper trail.

Because both agencies are arms of one state, this separation is a *governance* barrier, not a *technical* one. A state authority empowered to compel both can re-identify; the architecture does not prevent this, and we do not claim that it does. The institutional approximation of the distributed-de-anonymization ideal [Daza et al. 2022] is real and valuable, but it is an approximation, and we name its boundary rather than let it be discovered.

#### *4.2.2 Logging: per-agency records, not a combinable end-to-end audit*

We withdraw, explicitly, any claim that "every de-anonymization is gated and auditable." The implemented reality is weaker and we state the weaker truth. Each agency may log its own step — SIC logs which agent's reference was surrendered, under what cited authorization, and when; the MPS National Identity Service logs which reference was resolved and when. These are **two independent logs.** Crucially, no single party — including any oversight body — can today combine the two logs to reconstruct a complete "agent → identity" trace, because doing so would itself require holding both halves. The separation that delivers the privacy property also forecloses unified end-to-end audit. The system therefore provides **per-step accountability records but not combinable end-to-end auditability**, and we do not claim the latter — the most dangerous overclaim available in this design.

Stronger end-to-end auditability and stronger split-knowledge privacy are partly in conflict: making traces fully auditable by one body reintroduces a party who can correlate both halves. Section 4.4.3 discusses what oversight design could partially reconcile them — for instance, an oversight body empowered to demand and compare both logs under its own due-process constraints — framed as a needed reform, not a deployed feature [Solove 2011; Backes et al. 2014].

#### *4.2.3 Separation as two real agencies, and why it is institutional rather than cryptographic*

The separation of "who knows identity" from "who knows linkage" is realized by **two actual, organizationally distinct government agencies** — the MPS National Identity Service and SIC — not by a notional separation of roles within one operator. This is stronger than most escrow designs, which separate roles within a single trust domain. Neither agency can unilaterally re-identify: SIC holds linkage-without-identity, MPS holds identity-without-linkage, and recombination occurs only inside a legal authority that compels both. This is an institutional realization of the distributed-de-anonymization principle — a separation of powers applied to identity escrow through real inter-agency structure. We echo the honest limit: real-agency separation raises the bar but, both agencies being arms of one state, does not technically prevent a state empowered to compel both, and "two agencies" must not be read as "two adversarial jurisdictions."

We chose institutional separation rather than the cryptographic threshold mechanisms that the literature we cite already provides [Camenisch and Lysyanskaya 2001; Köpsell et al. 2006; Daza et al. 2022]. The honest answer is that the split is **inherited, not invented** — but we are precise about *what* is inherited. What we inherit is the MPS service's **anonymous-return mode**: the option, exercised at our choice, to receive a validity confirmation and a pairwise reference rather than the plaintext identity the service can also return to authorized government agencies. What we introduce is the **agent-binding construction** built on that mode — the SIC-held linkage half, the DID-to-agent-key verifiable credential, and the two-party trace protocol — which is what turns a bare anonymous authentication into *split-knowledge binding*. The MPS National Network Identity Authentication Public Service is designed — as government privacy policy for the national service — to return anonymous results to relying applications in its standard mode: a confirmation of validity plus a pairwise reference, with identity attributes not flowing to the relying party (plaintext identity is reserved for specifically authorized government agencies under law) [Measures for the Administration of the National Network Identity Authentication Public Service 2025]. SIC's agent-identity application integrates as an authorized relying party of that national infrastructure in this anonymous mode. The two-agency knowledge split is therefore the *deployed integration contract* of the MPS National Identity Service, not an escrow design improvised for this system. A cryptographic threshold scheme would require re-engineering the national authentication infrastructure itself, which is outside the agent layer's mandate; the anonymous-return mode is the sanctioned integration path. This matters because it inverts the natural suspicion that the structural variant was chosen for state convenience: the privacy split is the national infrastructure's own privacy-by-design baseline, which the agent layer composes with. We still concede plainly that, on abuse-resistance alone, a cryptographic threshold dominates; inheriting the structural split means inheriting its limits (Sections 4.2.1 and 4.3.1), and the migration question (Section 4.4) stays live — now with the honest caveat that migration would require change at the national-infrastructure level, not at the agent layer.

#### *4.2.4 Legal-process gating and subject non-notification*

Each disclosure is restricted by policy and legal obligation: SIC is required to surrender a reference, and the MPS National Identity Service to resolve it, only upon a valid legal instrument. We make the honest correction here that this gating is **policy-conditional, not technically enforced.** There is no technical control that refuses an authorized-but-improper request; the restraint is legal and procedural. We do not claim a technical gate — this matches most lawful-access regimes, and stating it plainly is more credible than implying a control that does not exist.

The subject of a de-anonymization is **not notified** when disclosure occurs within a legal proceeding, consistent with most states' lawful-access practice. Moreover, by design, neither agency even knows that a *person* was identified — SIC surrendered only a reference, and MPS resolved only a reference — so notification could come only from the legal authority, never from the system. We state this as a limitation, not a feature.

#### *4.2.5 Non-frameability*

An innocent principal cannot be framed for an agent action they did not authorize. At the protocol level, the binding is created through a cryptographically authenticated interaction with the MPS National Identity Service — verification, pairwise reference, and SIC-issued VC — so an attacker cannot forge a binding or substitute a different principal. We do not overclaim the audit chain: the two independent per-agency logs provide some basis for detecting irregularity, but, per Section 4.2.2, there is no single combinable end-to-end trail, so non-frameability rests primarily on the integrity of the binding step rather than on end-to-end audit [Daza et al. 2022; Backes et al. 2014]. We state it this way precisely because we have just disclaimed the combinable audit that a stronger claim would require.

### 4.3 Residual risks, not minimized

We now state what the constraints do *not* solve. We characterize these risks rather than minimize them.

#### *4.3.1 Authority over-reach or compromise*

The decisive residual risk is a state authority empowered to compel both agencies. Because the two agencies are arms of one state and the gating is policy-conditional (Section 4.2.4), nothing technical prevents lawful-but-improper recombination, nor bulk recombination if the legal framework permits it. The split-knowledge design defeats any *single* party but not a *coordinated state*. The Section 4.2 protections — real two-agency separation, two-step compulsion, per-agency logs — raise cost and leave traces but do *not* eliminate this risk, and we state this as the model's principal limitation rather than a footnote. Absent a combinable end-to-end audit (Section 4.2.2), the detection of systematic over-reach depends on external oversight that can demand both agencies' logs — a governance capability we discuss as a needed reform in Section 4.4.3, not as a deployed guarantee.

#### *4.3.2 Function creep*

A tracing authority initially authorized for serious crime may, over time, expand its tracing to administrative infractions, civil disputes, or routine surveillance — the well-documented pattern of surveillance function creep [Koops 2021; Solove 2011; Solove 2004; Richards 2015]. Legal-process gating helps only if the scope of admissible legal instruments remains narrow; if the legal framework expands, so does tracing. This is a governance risk, not a technical one, and it cannot be solved at the protocol layer; Section 4.4.3 answers it with institutional design intended to make creep detectable and reversible.

### *4.3.3 Collusion between tracing authority and business party*

If a tracing authority colludes with a platform or application — sharing de-anonymized data in exchange for metadata — the separation collapses. We declared this out of scope in Section 3.3.3, and we note it here as a genuine residual risk that the technical architecture cannot prevent. We do not pretend the boundary makes the risk disappear.

### *4.3.4 Straw-principal bindings*

A binding can be cryptographically genuine and socially fraudulent. An instigator recruits, pays, coerces, or identity-thefts a real person through the identity verification process, and lawful tracing then resolves — correctly, by the system's own lights — to the straw. The "definite legal principal" of Section 2.1 is definite but possibly not *responsible*. Every real-name registration regime has produced markets in verified accounts [Lee and Liu 2016; Thomas et al. 2013], and we assume this one will. Because identity verification is performed once per person rather than per agent (Section 3.2.1), the principal's marginal cost of binding additional agents after the one-time verification is low — the anti-straw protection rests entirely on the initial liveness and facial check. This is a genuine cost of the verify-once design relative to a per-agent re-verification regime: the straw need only pass identity verification once to bind an arbitrary number of agents. The per-agent linkage step (registering the agent's public key) confirms the principal's continuing consent but does not re-verify identity. We state the consequence for the accountability claim plainly: tracing reliably reaches *a* bound principal; reaching the *responsible* party can require a second, non-architectural step (Section 7).

### *4.3.5 Within-SIC cross-agent profiling*

Because the pairwise reference is stable per (person ↔ SIC), SIC can link every agent of one principal into a single pseudonymous profile — full cross-agent behavioral aggregation, one compelled disclosure away from a name. This is within the letter of the privacy property (Section 2.2) though not fully within its spirit. A design alternative exists: per-*binding* rather than per-*person* references would prevent this, at the cost of complicating reference-level revocation and multi-agent tracing. We name the trade-off; we do not resolve it [Solove 2011].

### *4.3.6 Degraded business-layer enforcement*

Section 2.1 calls transparency-, reputation-, and prevention-based accountability "complementary," but the architecture *degrades their substrate*. With no persistent cross-context identity, platforms cannot ban a principal — they re-bind new agents — reputation cannot accumulate across agents, and recidivism is invisible at the business layer [Gillespie 2018]. Escrowed anonymity does not merely decline to provide conceptions (i)–(iii); it weakens them.

There is an implemented partial repair: reference-level revocation is enforced at call time, because accountability-requiring calls carry a real-time-generated VP (Section 3.2.3), so gateways reject VP-bearing calls made on revoked credentials, and a principal's agents can be cut off at the *reference* level without identity disclosure. The business layer cannot ban, but the registry can, pseudonymously. This implemented power raises a new and honest question, however: revocation is governance-symmetric to tracing — it is a **pseudonymous kill switch** over a principal's agents. Its gating — who can order revocation, under what process, with what appeal and transparency — deserves the same institutional-design analysis that Sections 4.2 and 4.4.3 give tracing. We can state the governance partially. Blocklisting is initiated only by a formal request from authorized government agencies: SIC is technically capable of acting alone but is procedurally barred, and acts on submitted formal requests. The gate, like the tracing gate (Section 4.2.4), is policy and procedural, not technical — a symmetry we state explicitly. What remains to be settled, and what we do not claim to have settled, is which agencies qualify as authorized requesters, whether blocklisting bars only future bindings or also revokes existing credentials, what appeal or remedy a blocklisted principal has, and whether any transparency reporting exists.

#### *4.3.7 Cross-service correlation at the agent level*

The unlinkability property of Section 2.2 is identity-directed and person-level, and we scope it explicitly here, because the agent-level picture is different. An agent's identifier is public and globally stable across every service it touches; colluding services (adversary A2) can therefore join their retained VP records on the agent identifier and assemble a cross-service behavioral profile of that agent. This is by design — a stable public identifier is what makes agents addressable at all — and the profile is pseudonymous: it names an agent, never a principal, and no accumulation of business-layer records can cross the split-knowledge boundary to an identity. Two bounds on the pseudonymity deserve honest statement. First, pseudonymous profiles admit *behavioral* de-anonymization: an agent whose actions are self-revealing — booking one household's appointments, trading one firm's inventory — de-anonymizes its principal through content, not through the identity layer, and no identity architecture can prevent that. Second, whether a socially named agent exposes its principal's *other* agents at the business layer depends on whether VPs disclose the person-level DID: they do not. Services never see the pairwise reference, and selective disclosure withholds the DID, confirmed against the deployed presentation profile — so a socially de-anonymized agent does not, by that fact alone, expose the rest of its principal's fleet at the business layer. (Sub-agents sharing a parent's key pair are a separate exception: see Section 4.3.5 and L10 for the linkability that shared keys, not DID disclosure, create among them.) The property at its true strength: the architecture prevents identity-layer de-anonymization by business parties; A2's correlation succeeds at the agent level and fails at the person level.

#### *4.3.8 The composed objection*

The concessions of this section should also be faced in composition, because a critic will compose them: a biometrically verified registration system for AI agents, gated by policy rather than technology (Section 4.2.4), exempt from combinable end-to-end audit by its own design (Section 4.2.2), with no subject notification (Section 4.2.4), and equipped with a pseudonymous kill switch (Section 4.3.6) — in which the privacy property holds fully against business-layer adversaries and not at all against a state that compels both agencies (Section 4.3.1). On this reading, the design amounts to state legibility of the agent economy with privacy as its business-facing surface. We state that reading ourselves rather than leave it to be stated for us, and we answer what can be answered. Three elements of the composition are genuinely bounded: binding is voluntary under present law (Section 3.2.1); every business-layer party is structurally excluded from identity (Section 4.0); and each agency alone — including each of the two a state would compel — is defeated (Section 4.0). What the composed objection correctly identifies is that every *remaining* protection is institutional, and that the institutions are arms of one state. That is precisely the boundary Sections 4.2.1, 4.3.1, and 4.4.4 draw, and the reason this paper administers its own appropriateness test without adjudicating the result — a jurisdiction-specific legal judgment outside this paper's scope. The answer to the composed objection is therefore not a refutation but a jurisdictional condition: the design is defensible exactly where the tracing authority is credibly constrained, and nowhere else.

## 4.4 Administering the reflexive jurisdiction method

Section 2.5 defined the reflexive jurisdiction method. This section applies it: we state when escrowed accountability is appropriate, examine cross-border limits, identify the governance mechanisms the method surfaces, and administer the cost/takes analysis to the paper's own deployment jurisdiction.

#### *4.4.1 When is escrowed accountability appropriate?*

The proportionality framework of Section 2.4 supplies the conditions. Escrowed accountability is appropriate when (i) the relevant agent actions carry legal consequences — the domain the ex-post attribution thesis (Section 2.1) identifies as requiring attribution-based accountability; (ii) the dominant threat is private-sector exploitation rather than state over-reach, so that escrow's structural privacy removes identity from the parties most likely to abuse it; and (iii) the jurisdiction's judicial oversight credibly constrains the tracing authority.

The third condition is the framework's decisive boundary: where it fails, "do not deploy escrow here." The first condition — that the primary threat is private-sector — is an empirical and contested premise in any jurisdiction, and most acutely in our own. We do not assert it as a given. Section 4.4.4 faces it for our own case.

**Recourse for harmed third parties.** The ex-post attribution thesis has an honest cost. Because the architecture makes *private* de-anonymization impossible by design, a victim of an anonymous agent — a defrauded counterparty, a harmed bystander — has no path to redress except by triggering a lawful tracing process; they cannot themselves discover who is responsible. This is a deliberate, defended design choice. Subordinating private recourse to lawful process is the price of structural privacy [Landes and Posner 1975; Becker and Stigler 1974]. The defense is that private de-anonymization would also serve the adversaries the architecture is designed to defeat (A1, A2), and that the architecture preserves the tracing path while removing it from the business layer. A reader who finds this answer insufficient should reject the model, not the paper's candor in stating the cost.

#### *4.4.2 Cross-border limits*

Agent identities do not generalize across jurisdictions. An agent identity bound and verified under China's institutional structure is not automatically portable to a European eIDAS framework any more than a Chinese national ID card is valid as a travel document. The split-knowledge binding mechanism is implemented in *this* jurisdiction and bound to *these* agencies. Cross-border agent interaction would require mutual recognition agreements or a separate, jurisdiction-independent mechanism — neither of which currently exists for agent identities.

The proportionality framework (Section 2.4) makes a specific prediction about this: a jurisdiction with strong rule-of-law must not honor tracing requests from a jurisdiction without it, absent equivalent due process; as a consequence, the agent economy fragments at the rule-of-law boundary. This prediction is falsifiable — it will be confirmed or refuted by the actual pattern of mutual recognition agreements as agent-identity infrastructure matures — and it is the test by which we would have the framework judged.

#### *4.4.3 Governance against function creep*

The technical constraints of Section 4.2 are necessary but not sufficient. Preventing function creep — the drift from judicial tracing to routine administrative surveillance — requires institutional design, and we name three mechanisms: an independent oversight body with real enforcement power; mandatory periodic review and sunset provisions for the scope of tracing authorization; and public transparency reporting through aggregate tracing statistics [Anderson 2015; President's Review Group 2013]. The credibility of escrowed accountability depends entirely on the credibility of these governance mechanisms, which in turn depends on jurisdiction-specific institutional arrangements: the strength of judicial oversight over the tracing authority is the decisive variable, and the framework treats it as a question to be examined for each jurisdiction rather than assumed anywhere.

The same institutional-design demands apply to the implemented revocation power (Section 4.3.6). A pseudonymous kill switch over a principal's agents needs the same answers — who orders it, under what process, with what appeal, and with what transparency reporting — as tracing authorization does. We extend the oversight discussion to cover both powers symmetrically.

#### *4.4.4 Applying the method to this deployment*

The proportionality framework's decisive condition is credible judicial constraint, and a reader will apply that test to our own jurisdiction in a single step. We administer it ourselves rather than leave the inference to hostile hands.

We are direct about the part of the analysis we *can* settle. In the deployment jurisdiction — China — state authorities can lawfully compel both agencies; the architecture does not, and cannot, prevent this. What the system changes is the cost, the procedure, and the traceability of re-identification, not its possibility. We state this plainly, as a fact about this deployment and not as a hedged abstraction.

We are equally direct about the part of the analysis we do *not* settle. Whether the deployment jurisdiction's mechanisms of judicial oversight credibly constrain the tracing authority, and whether the dominant threat there is private-sector or state, are precisely the jurisdiction-specific institutional questions our framework exists to make explicit. The architecture determines what re-identification *costs*; the jurisdiction determines what it *takes*, and the latter is an institutional fact no architecture can supply. We therefore present the deployment as evidence of *feasibility at national scale*, and the framework as the instrument by which any jurisdiction — including this one — should be judged. The judgment itself requires a jurisdiction-specific legal analysis outside this paper's scope and the authors' expertise, so we make neither claim: not that the deployment satisfies the framework's conditions, and not that it fails them. Either claim would exceed our evidence and convert the case study into advocacy. What demonstrates the framework's analytical value is that it can interrogate the very system its authors built and surface exactly the question a jurisdiction-specific legal analysis would need to answer — independent of who ultimately renders that verdict.

---

## 5. Deployment Status and a Pre-Registered Measurement Plan

In this version of the paper, this section is a commitment rather than a report. The system is built and scheduled for public launch in Q3 2026; the deployment evidence that only its operators can provide does not yet exist in reportable form. Rather than publish placeholders, we do here what a measurement study can honestly do before launch: we state the method, the ethics constraints, the disclosure protocol, and the exact questions — including the unfavorable ones — that post-launch revisions of this paper intend to answer. Stating these in advance, in public, is itself a form of evidence discipline: it constrains the operators from selecting which measurements to disclose after seeing them. We use "pre-register" in the plain sense of stating questions and rules publicly before the answers exist — this timestamped arXiv version is itself the lodgment — not in the sense of a third-party registry deposit. We remain constrained, in a way we state openly, by the fact that some operational figures are government data that cannot be disclosed in an academic venue.

## 5.0 Method, ethics, and the intended dataset

**Method.** The lessons to be reported will be drawn from operational logs over the deployment period, incident and support records, and structured input from the operating team at SIC; where a lesson rests on a particular source, we will say so. We state the method explicitly because a reader's first question about deployment claims is "how do you know?", and lessons without provenance read as anecdote.

**Ethics and reporting.** Every figure we report will be aggregate; we will report no individual-level principal or agent data. The reporting itself must comply with the privacy properties the paper claims: we will perform no linkage that the architecture forbids, and in particular we will not combine any two figures in a way that would re-identify. We will note whether the log and record analysis underwent internal review under the operating institutions' governance, and we include the ethical-considerations and adverse-impact statements separately (see the *Ethics and Adverse-Impact Statement* at the end of the paper).

**Disclosure-tier protocol.** Because some figures are government data not disclosable academically, we classify every candidate figure before reporting it: **T1** already public; **T2** clearable through formal approval; **T3** publishable only as a band, an order of magnitude, or a normalized trend; **T4** qualitative lesson only; **T5** withheld. Banded reporting is legitimate and carries most of the evidentiary weight of exact counts — "more than $10^5$ agents," "fewer than $N$ tracing invocations" — and normalized trends (index = 100 at launch) show dynamics without disclosing absolute scale. Our designated T1 vehicle is public releases through the relevant government channels: where such a release publishes a figure (for example, agents bound in the first three months after launch), we will cite it with provenance and date, archive it, and use only the published value, verbatim; anything beyond the release falls back to T2–T5. We state the withholding policy openly: classes of data withheld under government-data rules will be named in the post-launch revision and in Section 7 (L11) rather than left as unexplained vagueness, because an explicit boundary is itself an honest data point about a state-operated system.

**Intended dataset, and a calibration rule.** Post-launch revisions aim to report, as data become available and clearable: (a) agent and principal counts and the time period; (b) binding-volume operational figures; (c) at least one candid failure or friction account; (d) the tracing-invocation count with an honest reading of it; and (e) whether per-action VPs and action-signing are used in practice. We set no schedule: operational data of this kind accrues — and clears government-data review — more slowly than paper revisions, and some items may arrive late or not clear at all (L11). What we fix in advance is the calibration rule: claims will be sized to the data actually reported. A defensible order of magnitude (T3) suffices for "infrastructure-scale"; if even that proves unpublishable, we will present the work as a deployment *experience report* of a national pilot rather than a deployment *measurement study*. The candid failure account is qualitative and is not excused by any disclosure constraint.

## 5.1 Launch status

The system is operated by a joint venture majority-owned by the State Information Center (SIC) in conjunction with the China Internet Network Information Center (CNNIC) as the national domain/address registry, and composes with the MPS National Network Identity Authentication Public Service for identity binding. Public launch is scheduled for Q3 2026. Scale and rollout figures will be reported in post-launch revisions under the disclosure-tier protocol of Section 5.0: public releases from relevant government channels cited verbatim where available, bands or normalized trends otherwise, and omissions named as such.

## 5.2 Pre-registered deployment questions

These are the observations a purely theoretical treatment cannot provide. We intend to answer each of the following in post-launch revisions, favorable or not, under the protocol of Section 5.0, as data become available and clearable.

### *5.2.1 Onboarding friction*

What was the hardest part of getting agents bound to principals through the national authentication system — user comprehension of the anonymous-authentication mode, integration friction with existing identity infrastructure, developer-adoption barriers, or something not on this list? We will report this candidly, including where it reflects a design decision that imposed cost on users, because the per-agent ceremony (Section 3.2.1) deliberately trades convenience for attribution integrity and we should be honest about the size of that trade.

### *5.2.2 Observed misuse and tracing invocation*

How many instances of agent misuse were observed over the deployment period (count or band), and how many times was the tracing mechanism invoked (count or band; possibly zero)? We commit in advance to the interpretive discipline this figure requires. A low or zero invocation count is genuinely ambiguous: it is consistent with genuine deterrence, with the immaturity of a young deployment, and with adoption too low to have yet attracted serious misuse. We will not read it as vindication. Even "zero invocations over period *T*" is a strong and publishable data point, but only if reported with that honest range of readings, and we bind ourselves to report it as such. This will be the empirical heart of the revised paper, and we decline in advance to let it carry more weight than it can bear.

### 5.3 Reproducibility statement

The system is a closed, state-operated national infrastructure tied to a national authentication service. It cannot be open-sourced or independently reproduced by external parties, and external verification is limited to the description we provide. We do not apologize for this; it is the reason the paper exists. The authorship — the system's architect and co-developer together with the operating institutions, SIC and CNNIC — gives the community something it cannot otherwise obtain: a first-hand, authoritative account of a real national system, including data only the operators can release. Deployment insight into a system that cannot be studied from the outside is precisely what this paper is structured to provide across its versions, and the contribution is the description, the pre-registered measurement protocol, and — in the post-launch revision — the data and analysis from inside the system, enabling the research community to examine, critique, and learn from a deployment that would otherwise be entirely opaque.

The flip side of insider authorship is real: the same position that makes the data available also creates an evaluative-independence concern, since we built and operate the thing we assess. We mitigate this in two ways that the reader can check. First, we commit to reporting unfavorable observations — failures, frictions, any misuse — with the same prominence as favorable ones, and the pre-registered questions of Section 5.2 are written to that standard, in public, before the answers are known. Second, we have framed Sections 4 and 7 as genuine self-critique rather than justification, including the residual risks we cannot solve and the reflexive jurisdiction method of Section 4.4.4. A reader should still weigh the source; we ask only that they weigh it against a paper that visibly resists boosterism.

---

## 6. Related Work

We organize related work by contrast rather than enumeration: every cited line of work is paired with an explicit statement of how our work differs. The cryptographic lineage (Section 6.4) is the section doing the real positioning work, and we keep it at length; the others we compress.

### 6.1 The dominant "anti-anonymous" agent-identity discourse

**Industry identity control planes.** A cluster of commercial platforms now offer "agentic identity control planes": end-to-end visibility, control, and governance for non-human identities, in which every agent receives a unique cryptographic identity, no agent is anonymous, every action is tied to a human, and the operator retains continuous auditability [Okta 2025; Ping Identity 2025; Descope 2025; authID 2025].

*Contrast:* these equate accountability with identifiability *and* with continuous visibility to the operator; ours withholds identity from the operator entirely, preserving accountability only through a due-process-gated path. The trust root differs as well: in these control planes the platform itself verifies and vouches for identity, so the binding is only as strong as a commercial operator's verification, whereas the binding here rests on government-performed verification of legal identity through the national authentication infrastructure. We are careful about the evidentiary status of these sources: they are vendor materials, and we cite them only as evidence of *industry practice and commercial direction*, never as evidence of a research or intellectual consensus. Where the paper needs to characterize what the field *believes*, we anchor that in peer-reviewed work [Chan et al. 2024]. The accurate phrasing is that industry practice overwhelmingly assumes no anonymous agents — not that "the field holds" this.

**Verifiable user–code binding.** A research line binds operator and agent code for accountability, using cryptographic attestation to prove who deployed which code: zkVM-based code-level authentication that treats the program binary as identity [BAID 2025, currently a preprint], OAuth2/OIDC extensions carrying agent-specific credentials and verifiable delegation chains [South et al. 2025], and a security architecture for governing agentic systems with cryptographic access-control tokens [SAGA 2025]. *Contrast:* we share the binding goal but add full business-layer anonymity and route disclosure through judicial escrow rather than verifiable public binding — *visible* accountability versus *latent* accountability. The identity substrate also differs: these schemes bind code to operator identities attested by platforms or by the schemes themselves, while our binding rests on government-verified legal identity.

**Public attestation / "Know Your Agent" reputation models.** A third line builds *public* identity, ownership, and reputation for agents so that counterparties can make trust decisions [Vouched 2025; Trulioo 2025; Birch and Hoffart 2025]. *Contrast:* these are deliberately public and reputation-bearing, while ours is non-public and anonymity-preserving, with accountability latent until lawfully invoked. The two optimize for different values — counterparty trust versus principal privacy subject to societal accountability.

## 6.2 Agent discovery, naming, and addressing

Several proposals provide naming, resolution, and capability discovery for agents: a DNS-inspired, PKI-based, protocol-agnostic agent directory currently in IETF Internet-Draft form [ANS 2025], and earlier work separating naming from capability mediation at scale [Sycara et al. 2001]. *Contrast:* none is operated by a national domain/address registry. In our system the agent identifier is issued and resolved by CNNIC, transposing an existing national registration authority to agents (Section 3.1) — an institutional arrangement we report as an operational fact of the deployment rather than claim as a contribution of this paper.

## 6.3 National and privacy-preserving digital identity for humans

The user-sovereign identity paradigm is closest to us in technique and farthest in trust model. eIDAS 2.0 and the EUDI Wallet provide user-held credentials with selective disclosure for EU citizens [Regulation (EU) 2024/1183], and the underlying DID/VC machinery is comprehensively surveyed [Mazzocca et al. 2025; Tan et al. 2023; W3C VC 2.0]. *Contrast:* these target natural persons and businesses, assume user-controlled disclosure, and provide no agent-oriented binding or addressing; we return to the trust-model difference analytically in Section 4.4. We note a further structural limitation, applying the paper's own standard of surfacing inconvenient facts about a design. eIDAS 2.0 Article 5a(16)(a) prohibits issuers — including Member State governments — from obtaining data that allows user transactions or behaviour to be tracked, linked, or correlated without explicit user authorisation. This is a legal prohibition against issuer-level surveillance, and it is the correct intent. But the current Architecture and Reference Framework relies on SD-JWT and mDOC attestation formats whose issuer signatures are static [Regulation (EU) 2024/1183]: the same signature value appears in every presentation of a given credential. An issuer who ever observes a presentation — through collusion with a relying party, or through a revocation check — can recognise its own signature and link the transaction to the user. The Wallet Unit Attestation introduces a persistent device identifier as a further linkability vector. In the terminology of Section 2.4: the user-sovereign model trusts a legal prohibition to prevent issuer tracking, but the technical architecture does not enforce what the law requires. The escrow model takes the opposite posture — it assumes the state can compel both agencies and asks what visibility that compulsion leaves behind (Section 4.2.1). The contrast is not that one model has privacy risks and the other does not; it is that they locate the risk differently, and the escrow model makes the compulsion path auditable. Separately, the MPS National Network Identity Authentication Public Service [Measures 2025] is not a comparator but the *substrate* our identity-association layer composes with: designed for human-to-application authentication with privacy-by-design anonymous return, it is the origin of our split-knowledge structure (Section 4.2.3), and we cite it here because omitting our own substrate from related work would be conspicuous.

## 6.4 Revocable / accountable anonymity: the cryptographic lineage

This is the section that must demonstrate command of the relevant literature and position the paper honestly within it. The property we pursue — anonymity to all parties except a lawfully authorized tracer — is the subject of a four-decade cryptographic research tradition that the agent-identity literature has largely overlooked.

**Chaum (1985): the foundational vision.** Chaum's "security without identification" argued that transaction systems could protect individuals from surveillance while still preventing abuse, rather than treating identification as the price of accountability [Chaum 1985]. *We inherit* the core conceptual move — accountability is not identification. *We add* the application to agents, the realization as national infrastructure, and the governance analysis.

**Identity escrow and the key-escrow taxonomy.** The generic concept — identity revealed only to a designated escrow agent under defined conditions — carries an established name: *identity escrow* [Kilian and Petrank 1998], and escrowed-access systems were taxonomized during the key-escrow debate [Denning and Branstad 1996]. *We inherit* the concept and, for the generic property, its name. *We add* the institutional locus of the split: where identity-escrow constructions divide capability across cryptographic escrow agents, our split lies between two real agencies' knowledge stores — the sense in which "split-knowledge binding" (Section 2.2) names an institutional, not cryptographic, realization.

**von Ahn et al. (2006): selective traceability.** von Ahn, Bortz, Hopper, and O'Neill formalized *selective traceability* — a sender can be traced only with the cooperation of an explicitly specified set of parties — and *noncoercibility* [von Ahn et al. 2006]. *We inherit* the formal property of selective traceability. *We add* a national-scale instantiation in which the set of parties is institutional roles rather than cryptographic keys.

**Köpsell et al. (2006): the closest prior match.** Köpsell, Wendolsky, and Federrath proposed *revocable anonymity* based on threshold group signatures and threshold atomic proxy re-encryption, motivated by the EU Data Retention Directive; in their scheme normal activity remains anonymous, malicious activity is traceable, and the user's identity is revealed *to no entity involved in the revocation procedure except the law-enforcement agency* [Köpsell et al. 2006]. This is the closest prior conceptual match to the property our system provides — the phrase "revealed to no entity except the law-enforcement agency" is nearly word-for-word our design property. *We inherit* the core property: revocable anonymity with law-enforcement-only disclosure. *We add* agent-orientation rather than human communication, realization as national infrastructure, and the governance framework of Sections 4.2–4.4.

**Camenisch and Lysyanskaya (2001): anonymous credentials with optional revocation.** CL01 introduced anonymous credentials with optional anonymity revocation, in which an anonymity-revocation manager can recover a user's identity but is trusted not to exercise that power except when required, and noted that this trusted role "can be implemented in a distributed fashion to weaken the trust assumptions" [Camenisch and Lysyanskaya 2001]. *We inherit* the anonymous-credential primitive and the distributed-revoker recommendation. *We add* an *institutional* realization of that two-decade-old recommendation: split-knowledge binding across two real agencies — identity-without-linkage versus linkage-without-identity, recombined only by a compelling legal authority — achieved by inter-agency structure rather than a new cryptographic mechanism, and we are explicit that this is structural, not a cryptographic threshold.

**The Mutual Accountability Layer (2022): the critique formalized.** The Mutual Accountability Layer formalizes the hazard that a de-anonymization manager can abuse its power by improperly revoking anonymity without the user's awareness, and defines *mutual accountability*: both accountability (the user can be held responsible) and non-frameability (an innocent party cannot be falsely accused) must be provably achieved for both user and manager [Daza et al. 2022]. *We inherit* the formal adversary model — the malicious manager — and the non-frameability requirement. *We add* that this is precisely the adversary our Section 3.3 centers and our Section 4 must answer to; we treat non-frameability and constrained de-anonymization as the standard against which our design is measured.

**BackRef (2014).** BackRef provides verifiable backward accountability in anonymous communication networks with formal no-false-accusation guarantees [Backes et al. 2014]. *We inherit* its formal treatment of non-frameability in anonymous systems.

**The exceptional-access critique: Keys Under Doormats.** The policy-side lineage is as canonical as the cryptographic one. Abelson et al.'s 1997 and 2015 reports state the technical community's standing case against exceptional-access mechanisms — concentrated attack targets, added complexity, and jurisdictional precedent [Abelson et al. 1997; Abelson et al. 2015]. An escrow paper that omitted them would invite the inference of evasion; we cite them and answer them (Section 4.1.1): the split-knowledge design answers the concentrated-target argument, the institutional-rather-than-protocol mechanism partially blunts the complexity argument, and the precedent argument is faced, not answered, in Section 4.4.4. *We inherit* the systemic-risk framing and the burden of proof it assigns to any escrow design. *We add* an institutionally distributed design with no single recovery point — partially responsive, and honestly bounded.

**Positioning.** We introduce no new cryptographic primitive. Our contribution is an applied, national-scale, agent-oriented realization within this well-studied family, inheriting both its core mechanism and its central critique. Stating this lineage candidly is not a concession but a requirement: it is what distinguishes a defensible escrowed-anonymity claim from a naïve one.

---

## 7. Limitations

We state the limitations briefly, honestly, and without defensiveness, pointing to where each is treated in detail.

- **L1 — Conditional anonymity; state-level re-identification not technically prevented.** Split-knowledge binding defeats any single party but not a state that compels both agencies; gating is policy-conditional, with no combinable end-to-end audit. (Sections 2.2, 4.2, 4.3.1.)

- **L2 — Non-reproducibility.** The system cannot be open-sourced or independently reproduced; external verification is limited to the description provided. (Section 5.3.)
- **L3 — Metadata and traffic-analysis excluded.** Network-layer correlation attacks are out of scope. (Section 3.3.3.)
- **L4 — Jurisdictional boundedness.** The model depends on a specific national institutional and legal context; generalizability to jurisdictions with different judicial-oversight regimes is untested and likely bounded. (Section 4.4.2.)
- **L5 — Deployment evidence deferred.** This version precedes the system's public launch (scheduled Q3 2026) and reports no operational measurements; Section 5 pre-registers what will be reported and how. Even once reported, the evidence will be from a single system in a single jurisdiction over a finite period, and claims about deterrence, misuse patterns, and long-term governance effects cannot be extrapolated. (Section 5.)
- **L6 — Evaluative independence.** The authors include the system's architect and co-developer (Red Date Technology) and its operating institutions (SIC, CNNIC), and the first author holds a commercial interest in the joint venture that operates the platform. This insider position is what makes the data available, but it limits evaluative independence; we mitigate by reporting unfavorable observations with equal prominence and inviting external scrutiny, but readers should weigh the source. (Section 5.3; Ethics statement.)
- **L7 — Domain scope of the accountability conception.** We argue the ex-post attributive conception is appropriate for legally consequential agent actions; we do not claim it suffices for domains where transparency-, reputation-, or prevention-based accountability is the dominant need. (Section 2.1.)
- **L8 — The framework's conditions are not adjudicated for the deployment jurisdiction.** The paper supplies the appropriateness test and administers it reflexively, but does not perform the jurisdiction-specific legal analysis an actual verdict requires; the deployment must not be read as evidence of normative appropriateness. (Section 4.4.4.)
- **L9 — Straw-principal bindings.** Tracing reaches a definite bound principal, who may be a recruited or coerced stand-in; reaching the responsible instigator can require non-architectural investigation. (Section 4.3.4.)
- **L10 — Sub-agent delegation scope; shared-key attribution; within-registry linkability.** Sub-agents that share a parent agent's MCP server inherit the parent's linkage and key pair without a separate linkage step (Section 3.2.1). Because the key pair is shared, signed actions and VPs cannot distinguish a sub-agent from its parent; attribution resolves only to the shared-key cluster, not to the individual sub-agent that acted — and the shared credential also makes the cluster's calls linkable to one another at the business layer, independent of the DID-disclosure question addressed in Section 4.3.7. Agents with independent MCP servers require their own linkage. All of one principal's agents share a stable pairwise reference, linkable by SIC into a pseudonymous profile. (Sections 3.2.1, 4.3.5.)

- **L11 — Disclosure constraints on deployment data.** Some operational figures involve government data not disclosable in an academic venue; in the post-launch revision, affected values will be reported as bands or normalized trends, or omitted, with the withheld classes named. Readers should weigh Section 5's evidence accordingly. (Section 5.0.)
- **L12 — Biometric gate and verification-failure equity.** Binding passes through biometric verification (liveness, facial) at the MPS service; principals whom verification fails are excluded from binding, and recourse paths are a property of the national authentication service outside this paper's scope. The demographic skew documented for face-recognition systems [Buolamwini and Gebru 2018] makes this an equity-relevant limitation. (Section 3.2.1; Ethics statement.)

---

## 8. Conclusion

This paper has made five contributions. **Split-knowledge binding** (Section 2.2) — a coined mechanism for escrowed accountability, differentiated from prior cryptographic constructions by its institutional rather than cryptographic separation. **The ex-post attribution thesis** (Section 2.1) — the argued claim that for AI agent actions with legal consequences, only attribution-based accountability carries legal force. **The accountability surface** (Section 2.3) — a design concept identifying which agent actions leave identity-bearing traces, with a mapped granularity spectrum and trade-off. **A proportionality framework for identity escrow** (Section 2.4) — a decision structure grounded in legal proportionality that selects among three trust architectures and predicts fragmentation at the rule-of-law boundary. **The reflexive jurisdiction method** (Section 2.5) — an evaluative standard, demonstrated by administering the framework to the paper's own deployment in Section 4.4.

These contributions are instantiated in a national agent-identity system built for public launch in Q3 2026. The system is evidence of feasibility at national scale; the framework is the instrument by which any deployment — including this one — should be judged. We have pre-registered the protocol under which deployment evidence will be reported (Section 5), stated precisely the conditions under which escrowed accountability is and is not appropriate (Section 4.4), and been explicit about the residual risks the architecture does not solve (Section 4.3, Section 7).

As agents proliferate, the choice between "no anonymous agents" and "accountable anonymity" is not, at bottom, technical; it is political and institutional. This paper offers conceptual tools for making that choice explicitly rather than defaulting to identifiability. The design point is real, it has been realized, and it is now available to be analyzed — which is all we claim for it.

---

## Ethics and Adverse-Impact Statement

This work reports on a national identity system for AI agents in China — built and scheduled for public launch in Q3 2026 — and is, by its subject matter, ethically consequential. We address the principal considerations directly.

**On the system's dual-use character.** The architecture concentrates a re-identification capability in the state, gated only by policy and legal process rather than by technology (Sections 4.2.4, 4.3.1). We do not present this capability as benign, and we have administered our own appropriateness framework to the deployment without adjudicating whether it passes — a jurisdiction-specific legal judgment outside this paper's scope (Section 4.4.4). The paper's purpose is to make the trade-off analyzable, not to advocate adoption.

**On the data to be reported.** This pre-launch version reports no operational figures. In the post-launch revision, all figures will be aggregate; no individual-level principal or agent data will be reported, and the reporting will perform no linkage the architecture forbids (Section 5.0). Some operational figures are government data not disclosable in an academic venue; these will be reported as bands or normalized trends, or withheld, with the withheld classes named (Section 5.0, L11). Publicly released figures will be cited to their public source.

**On biometric verification.** Binding passes through identity verification at the MPS service, including liveness and facial checks (Section 3.2.1) — a state biometric processing event whose governance belongs to the national authentication infrastructure, not to the agent layer, but whose cost we count as part of this system's total privacy footprint: it occurs once per person, and no biometric data reaches the agent-identity layer (SIC receives only the pairwise reference). The equity consequence also belongs in this statement: principals whom facial verification fails — a failure mode with documented demographic skew in face-recognition systems [Buolamwini and Gebru 2018] — are excluded from binding and therefore from whatever the credential gates; verification-failure recourse is a property of the MPS service that we cannot specify here, and we flag it as an equity-relevant unknown (L12).

**On adverse impacts.** The most serious adverse impact is that an architecture designed to protect principal privacy from the business layer also builds infrastructure whose state-level re-identification and pseudonymous-revocation powers could, under weak oversight, be exercised outside the lawful-process constraints the architecture assumes (Sections 4.3.1, 4.3.2, 4.3.6, 4.4.3). We have surfaced these risks rather than minimized them, and we have argued that the credibility of the model depends entirely on governance mechanisms whose adequacy is a jurisdiction-specific question the paper does not resolve.

---

## Competing Interests

**On evaluative independence.** The authors built and operate the system they assess (Section 5.3, L6). We have committed to reporting unfavorable observations with equal prominence and have framed the analysis as self-critique; readers should nonetheless weigh the source.

**On competing interests.** The first author is the founder and CEO of Red Date Technology (Hong Kong) Limited, a co-developer of the platform described in this paper and a shareholder of the joint venture that operates the platform under SIC. The remaining authors are affiliated with the operating institutions (SIC, CNNIC) and with the China Organization Data Service of the State Administration for Market Regulation. We state this plainly: this is an insider account with both institutional and commercial interests in the system it assesses, and the mitigations of Section 5.3 and the caution of L6 should be read with that weight.

---

## AI-Usage Disclosure

AI assistance tools were used in preparing this manuscript for literature search, drafting, editing, consistency checking, and citation verification. All AI-assisted passages were reviewed, edited, and verified by the authors. No deployment figure or empirical observation was generated by AI. Generative AI tools are not listed as authors, and the authors take full responsibility for the entire content of this article.

---